\documentclass{desyprocA4}

\begin{document}
\title{Determination of Higgs-boson couplings (SFitter)}

\author{{\slshape Michael Rauch\footnote{for the SFitter collaboration}}\\[1ex]
Institut f\"ur Theoretische Physik, Karlsruher Institut f\"ur Technologie (KIT), Karlsruhe, Germany}

\contribID{xy}

\confID{1964}  
\desyproc{DESY-PROC-2010-01}
\acronym{PLHC2010} 
\doi  

\maketitle

\begin{abstract}
After the discovery of a Higgs boson, the next step is to measure its
properties and test their accordance with the predictions of the
Standard Model, in particular the couplings of the Higgs boson.
In this talk we discuss what information the LHC will be able to give us
over the coming years, and what remains as a task for a future Linear
Collider.

Using the well-established SFitter framework, we map measurements onto a
weak-scale effective theory with general Higgs boson couplings. Our
sophisticated error treatment allows us to take all theory and
experimental errors, including arbitrary correlations, fully into account.
\end{abstract}

\section{Introduction}

Completing our understanding of the electro-weak symmetry-breaking
mechanism is one of the main tasks for present and future particle
colliders. In the Standard Model (SM), this is accomplished by
introducing a complex $SU(2)$ doublet, the Higgs field, which obtains a
vacuum expectation value (vev)~\cite{higgs,reviews,HiggsXSWG}. Three of
the four degrees of freedom form the longitudinal modes of $W$ and $Z$
bosons, while the remaining one becomes a physical particle, the Higgs
boson.  Interactions between these gauge bosons and the Higgs field are
introduced automatically via the latter's kinetic term, while
interactions with fermions are added via Yukawa-type couplings.
Replacing the Higgs field by its vev then yields mass terms for the
gauge bosons and fermions. Therefore, the couplings of the Higgs boson
to the other particles are fixed and proportional to the measured masses
and the vev.

The mass of the Higgs boson is the only remaining unknown parameter in
the SM. Direct searches by LEP~\cite{Collaboration:2008ub},
Tevatron~\cite{HiggsTevatron} and in particular the LHC experiments
ATLAS~\cite{HiggsATLAS} and CMS~\cite{HiggsCMS} have excluded large
parts of the parameter space, leaving only a small window around 125
GeV. High-mass values, where the experimental sensitivity drops again,
are strongly disfavoured by indirect constraints from electro-weak
precision data~\cite{ZFitter, Flacher:2008zq}. 
As mentioned before, the Higgs couplings in the SM are completely
determined by the known particle masses. Therefore, we can use these
theoretically predicted values and compare them to future measurements
of Higgs boson
channels~\cite{duehrssen,Lafaye:2009vr,Bock:2010nz,Bonnet:2011yx,HiggsFits2012,SFitter2012,HiggsILC}.
Thereby, we assume that the discrete quantum numbers, like its CP
property or spin~\cite{wbf_vertex}, are identical to the SM expectation.
Many models of new physics predict deviations in the Higgs couplings,
which can then be measured. Examples include models with an extended
Higgs sector, like the two-Higgs doublet structure e.g.\ in
supersymmetry~\cite{susy}, or also Higgs portal
models~\cite{Patt:2006fw}, but modifications can also be more elementary
as in composite models~\cite{Espinosa:2010vn}, where the Higgs boson
emerges as a pseudo-Goldstone boson of a new strongly-interacting sector.

A correct treatment of all errors is important to obtain correct
results. As in the Higgs boson channels rates are measured, these
statistical errors are of the Poisson type. Additionally, there are
systematic errors, which are correlated, and we implement the full
correlation matrix between different measurements. Theory errors are
best described as box-shaped~\cite{HiggsXSWG}, using the prescription
of the RFit scheme~\cite{ckmfitter}. In the SFitter tool~\cite{sfitter}
these different types of errors are fully implemented. As output we
obtain a fully-dimensional log-likelihood map, which we can then reduce
to plotable one- or two-dimensional distributions via both Bayesian
(marginalisation) and Frequentist (profile likelihood) techniques.
Furthermore, a list of best-fitting points is obtained.

\section{Setup of the Calculation}

As the underlying model of our study we assume the Standard Model with a
generalised Higgs sector, where the Higgs couplings can take arbitrary
values. These are parametrised in the following way: Couplings to
particles $i$, which are present at tree-level in the SM, are modified
according to
\begin{equation}
g_{iiH} \rightarrow g_{iiH}^{\text{SM}} ( 1 + \Delta_{iiH} ) \ .
\end{equation}
As a global sign flip of all couplings is not observable, we always take
$g_{WWH}$ to be positive, \textit{i.e.} $\Delta_{WWH} > -1$.
Additionally, there are two important loop-induced couplings present,
namely those to gluons and photons. They are altered in the following
way:
\begin{equation} 
g_{iiH} \rightarrow g_{iiH}^{\text{SM}} ( 1 + \Delta_{iiH}^{\text{SM}} +
\Delta_{iiH}) \ .
\end{equation} 
These can receive two types of contributions. First, there are
contributions from changing the tree-level couplings, 
$\Delta_{iiH}^{\text{SM}}$. Second, there can be additional
dimension-five contributions $\Delta_{iiH}$. They originate from new
particles running in the loop, e.g.\ the supersymmetric partners in SUSY
models. 
The numerical values of the couplings are obtained from a modified
version of HDecay~\cite{Djouadi:1997yw}.
Also the masses of the Higgs boson and the top and bottom quark
are added as free parameters and corresponding measurements constrain
them to their experimentally measured value. 
Additionally, we define $\Delta_H$ as a single free parameter that
changes all (tree-level) couplings simultaneously.

The total width of the Higgs boson is too small to be measured directly
at the LHC. Therefore we have to make one single model assumption about
how to treat the total width, which we take as
\begin{equation*}
\Gamma_\text{tot} = \sum_\text{obs} \; \Gamma_i(g_{iiH}) 
+ \text{generation universality} \ .
\end{equation*}
This means that there are no further contributions from Higgs decays
into invisible particles.  The assumption about generation universality
is important as the Higgs has a significant branching ratio of several
percent into unobservable particles (e.g.\ charm quarks) for which at
the LHC there is no possibility to measure them, and neglecting them
would introduce a bias.
Further details of the setup have been described in
Refs.~\cite{Lafaye:2009vr, Bock:2010nz}.
We will not consider any couplings that can only be measured with very
high luminosity or not at all. This includes the only second-generation
Yukawa coupling that might be measurable at the LHC, namely those to
muons~\cite{wbf_mu,tth_mu}, as well as the Higgs
self-couplings~\cite{selfcoup, higgs_pot, quartic, selfilc}.

\section{Results}
\subsection{Expectations for the LHC at 14 TeV}

The measurements that enter our analysis are derived from an ATLAS
Monte Carlo study performed for an integrated luminosity of 30 fb$^{-1}$
and assuming a centre-of-mass energy of 14
TeV~\cite{Lafaye:2009vr,duehrssennote}. We perform a simulation with
typically 5000 toy Monte Carlos, where we smear the signal and
background expectations according to their corresponding errors, and
fit the resulting Higgs couplings.

\begin{figure}
\centerline{
\includegraphics[width=0.45\textwidth]{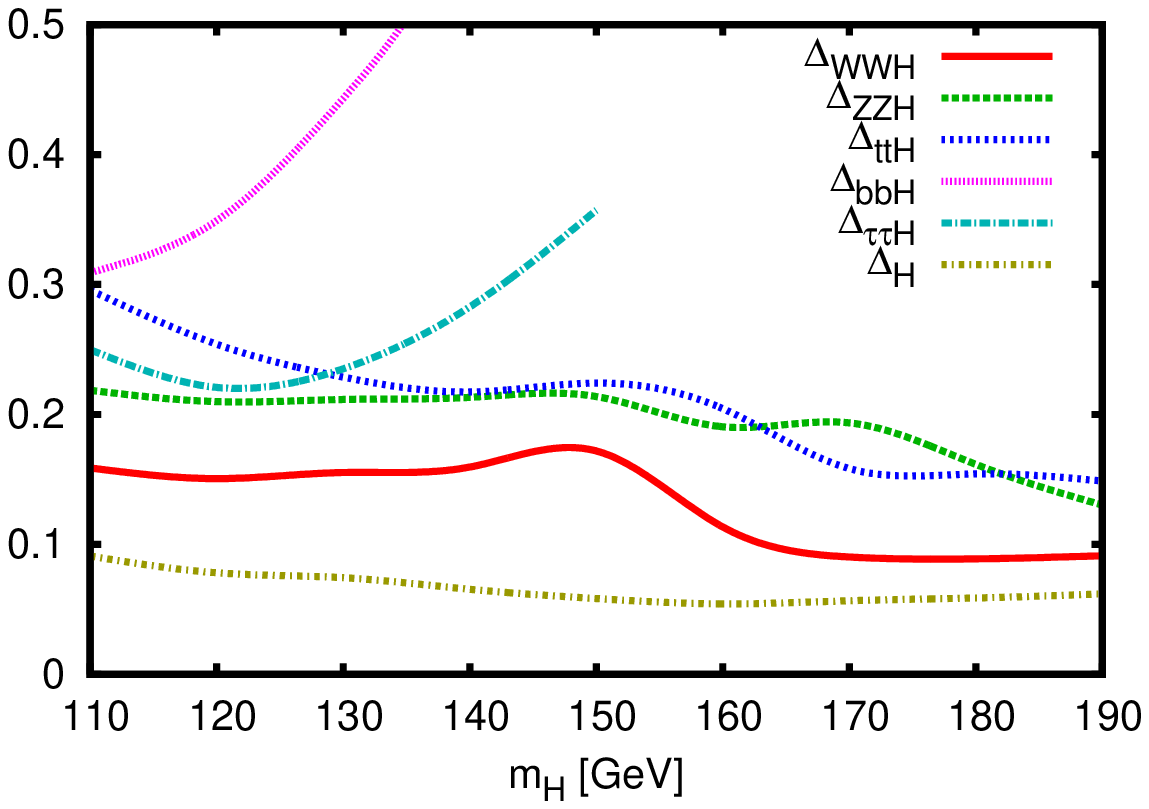}
\quad
\includegraphics[width=0.45\textwidth]{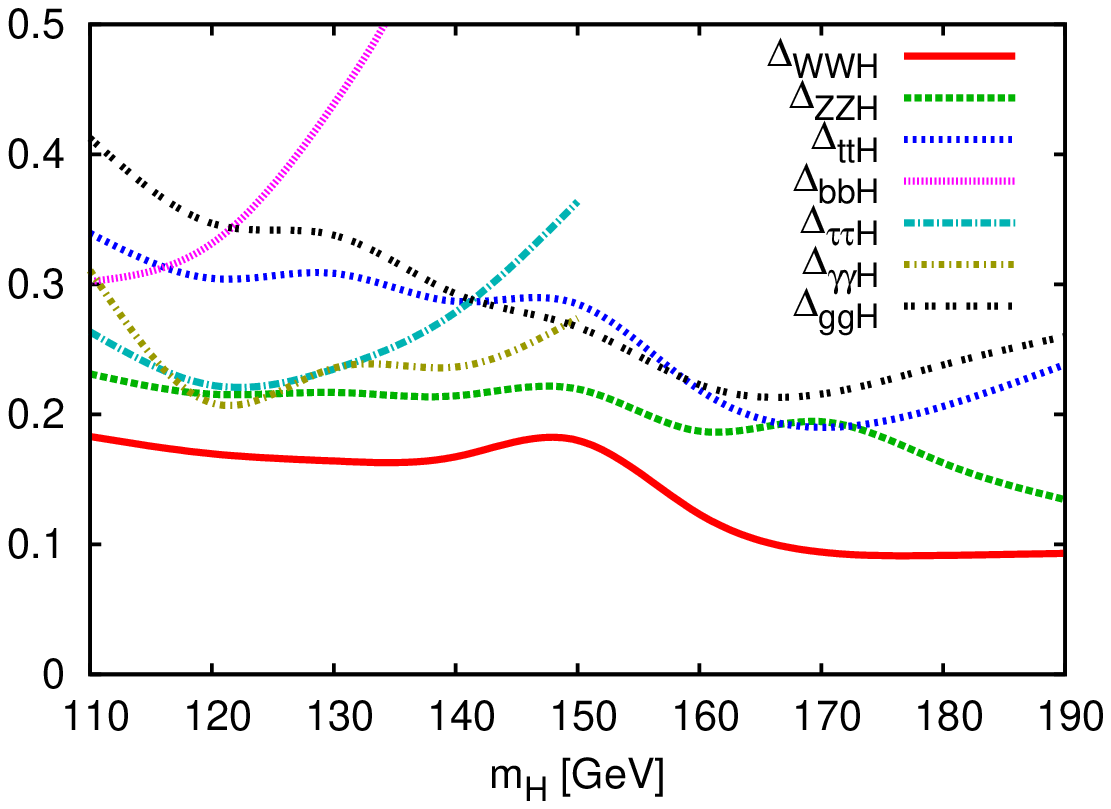}
}
\caption{Error on the Higgs-boson couplings as a function of the Higgs
mass without \textit{(left)} and including \textit{(right)} additional
dimension-five operators. The left-hand plot also includes the result
for a single-parameter modification $\Delta_H$.
Results are for the LHC at 14 TeV centre-of-mass energy and an
integrated luminosity of 30 fb$^{-1}$, assuming SM Higgs couplings.
Figures taken from Ref.~\cite{Rauch:2011sa}.}
\label{fig:mh14}
\end{figure}

In Fig.~\ref{fig:mh14} we show the results of our analysis. The
different curves denote the 68\% CL errors on the $\Delta_{iiH}$
parameter for the respective coupling. As input value for the signal
strength we assume a SM Higgs boson of the given mass value and note
that for reduced couplings the change in the absolute value of the
errors is small. On the left-hand side of the figure we present results
where additional contributions from dimension-five operators have been
neglected. Also shown is the result for the single-parameter modifier
$\Delta_H$. On the right-hand side the dimension-five operators are taken
into account as well.
In both cases the coupling of the Higgs to $W$ bosons can be measured best,
between 10 and 20\% over the whole mass range. The dimension-five
operators thereby reduce the sensitivity to this coupling somewhat.
Yukawa couplings to bottom quarks and $\tau$ leptons can only be
determined with good accuracy for Higgs masses below 140 to 150 GeV, as
for larger masses the corresponding branching ratios become too small. 
The top quark is strongly affected by the dimension-five operators
Without these operators the gluon-fusion production processes
contribute to the precision of this coupling. Including them, the
top-quark coupling needs to be determined by the badly measurable
top-quark-associated production modes, and gluon-fusion production then
pins down the size of the additional operators relative to the top quark
coupling.

\subsection{Extrapolation to 7 TeV}

To get an estimate of what to expect from the LHC in the near future, we
have extrapolated these studies to a centre-of-mass energy of 7 TeV.
For the backgrounds the inclusive cross sections of the
individual contributions were computed with
Sherpa~\cite{Gleisberg:2008ta} and the event rate scaled according to
the numbers obtained.  For the signal we assume that the signal
efficiencies, i.e.\ the number of signal events remaining after the
selection cuts and detector acceptance corrections relative to the
original rate, stays unchanged.  The cross sections themselves for both
centre-of-mass energies are taken from Ref.~\cite{HiggsXSWG}.  As the
expected precision on the couplings will be rather low, only the case of
vanishing additional dimension-five operators is considered here. 

\begin{figure}
\centerline{
\includegraphics[width=0.45\textwidth]{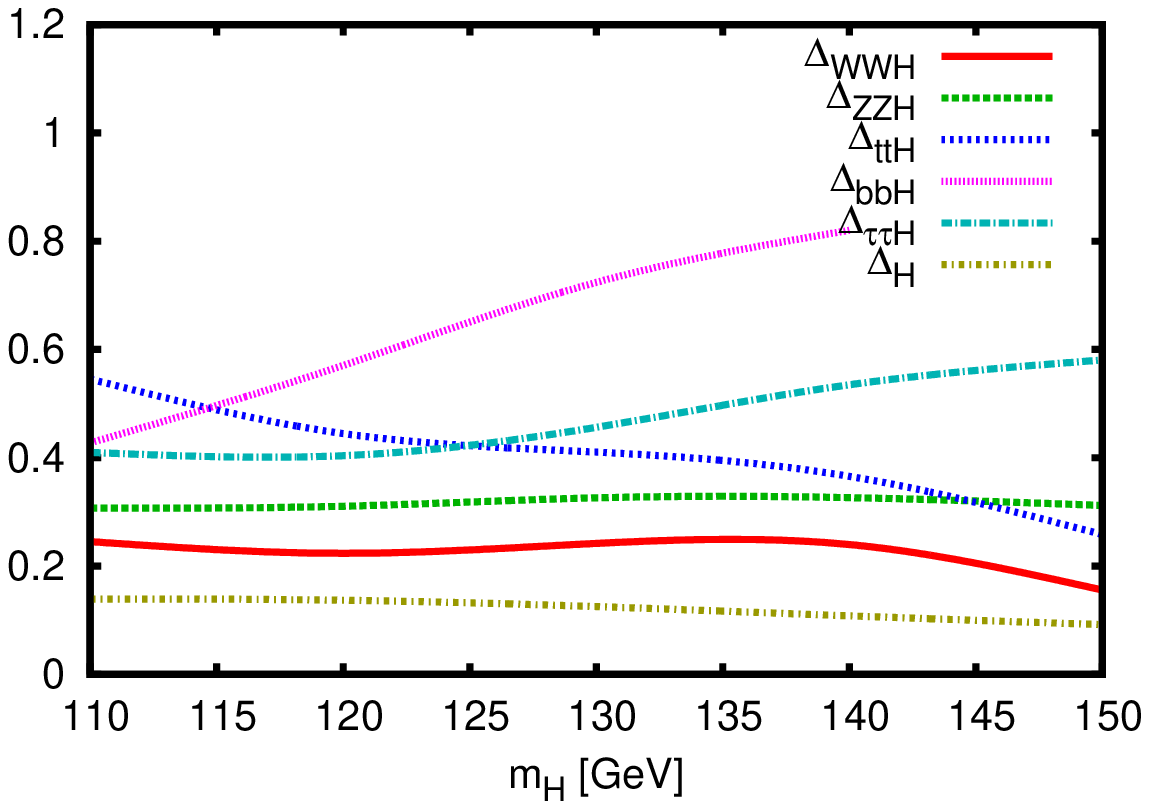}
\quad
\includegraphics[width=0.45\textwidth]{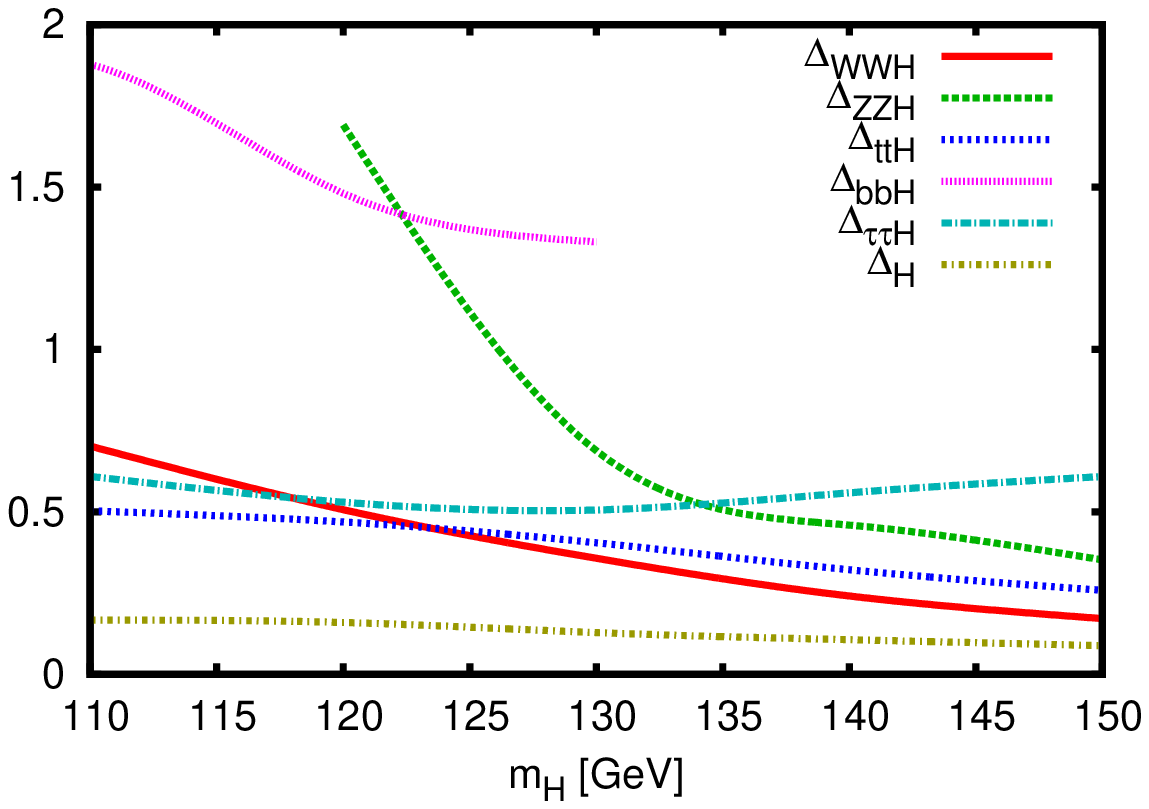}
}
\caption{Error on the Higgs-boson couplings as a function of the Higgs
mass including \textit{(left)} and without \textit{(right)} the subjet
analyses in the Higgsstrahlung production processes with decays into
bottom quarks.
Results are for the LHC at 7 TeV centre-of-mass energy and an
integrated luminosity of 20 fb$^{-1}$, assuming SM Higgs couplings,
obtained by extrapolating the 14-TeV-Monte-Carlo studies. }
\label{fig:mh7}
\end{figure}

In Fig.~\ref{fig:mh7} we show the corresponding results for an
integrated luminosity of 20 fb$^{-1}$, corresponding to approximately
what is expected for the end of 2012\footnote{The increased
centre-of-mass energy of 8 TeV for the 2012 run can be approximated by a
corresponding increase in the integrated luminosity.}.  
On the left-hand side we include all channels of the 14-TeV analysis.
We observe the same principal behaviour as in Fig.~\ref{fig:mh14}, but
with a significant increase in the expected errors. Nevertheless,
with this amount of data a determination of $\Delta_H$ with a
precision of 14\% is already possible for a Higgs boson mass of 125 GeV. 
On the right-hand side the channels making use of subjet
techniques~\cite{subjet} are removed. These consist of Higgs bosons
produced in association with a $W$ or $Z$ boson, where the Higgs decays
into bottom quarks and the decay products are required to be strongly
boosted in order to reduce backgrounds. A significant drop in accuracy
can be observed mainly for two couplings. The coupling to $Z$ bosons is now
predominantly determined by the decay of the Higgs to four leptons, which
suffers from low event numbers for lighter Higgs masses. The
bottom-quark Yukawa coupling has to rely on the top-quark-associated
production channel with decays into bottom quarks as well as its
contribution to gluon-fusion production. The first one suffers from
a large combinatorial background, while in the second case the
bottom-quark loop is only a small contribution. The badly determined
bottom coupling then influences all other couplings via the total width.

\subsection{Results and Expectations from Current Measurements}

\begin{figure}
\centerline{
\includegraphics[width=0.45\textwidth]{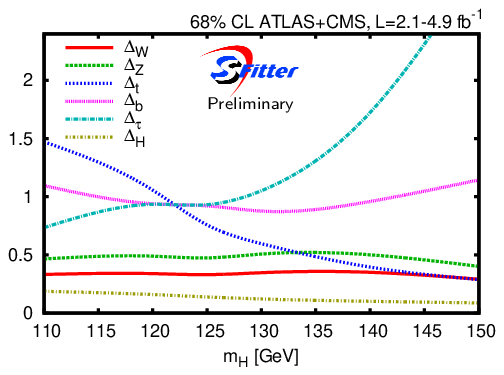}
\quad
\includegraphics[width=0.45\textwidth]{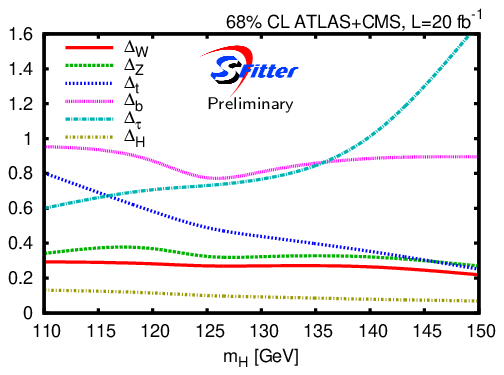}
} 
\caption{Precision of tree-level Higgs-boson couplings as well as
single-parameter modifier as a function of the Higgs mass using current
LHC searches as input. Results are presented for the luminosities used
in the analyses \textit{(left)} and extrapolated to 20 fb$^{-1}$
\textit{(right)}. Figure from Ref.~\cite{SFitter2012}.}
\label{fig:curr}
\end{figure}
\begin{figure}
\centerline{
\includegraphics[width=0.45\textwidth]{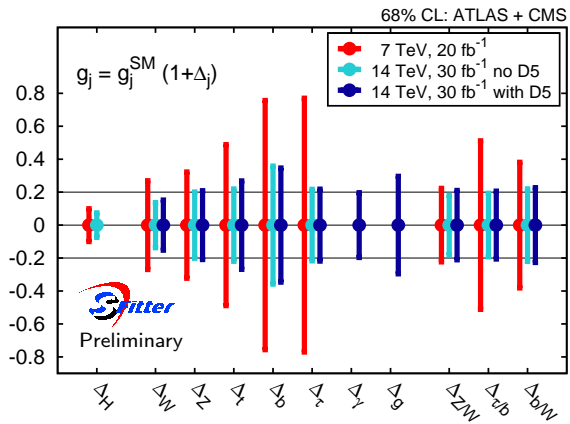}
} 
\caption{Overview of LHC expectations in different scenarios for energy
and integrated luminosity for a SM Higgs boson with mass of 125 GeV. Figure
from Ref.~\cite{SFitter2012}.}
\label{fig:summary}
\end{figure}

With direct search results available from the LHC, we can update the
results of the previous subsection using the actual background
expectations and errors as described in the analyses~\cite{HiggsATLAS,
HiggsCMS}. Thereby we assume as input that there is a SM Higgs boson at
the considered mass value and add a SM Higgs signal to the background
expectations. These results are depicted in Fig.~\ref{fig:curr}. On the
left we show errors on the Higgs couplings using for each measurement
the luminosity for which the analysis has been performed.  With this
data a precision of 14\% on $\Delta_H$ is already possible, and the
couplings to the weak bosons can also be measured fairly precisely. The
error on the top-quark Yukawa coupling is mostly determined by Higgs
production via gluon-fusion decaying into a $W$ pair. The sensitivity of
this channel drops rapidly below 125 GeV, leading to the observed
behaviour of the top-quark coupling. On the right-hand side we present
expectations when extrapolating all analyses to an integrated luminosity
of 20 fb$^{-1}$. This extrapolation is done blindly, i.e.\ the
improvement is purely statistical. The precision on the single-parameter
modifier now reaches 9\% for a Higgs mass around 125 GeV. 

In Fig.~\ref{fig:summary} the different results shown previously are
summarised for a hypothetical Higgs boson at 125 GeV assuming SM
couplings as a central value. The three values on the right-hand side of
the plot show errors on ratios of couplings. While for the $Z$ over $W$
Higgs couplings at 7 TeV only a small improvement over the absolute
measurements is achievable, the situation is different for the two other
ratios involving the bottom Yukawa coupling. Here correlations are
important and therefore the ratio is better determined. At a 14 TeV LHC
the situation is different. Using ratios yields no improvement over
absolute values in any case.

\section{Models of New Physics}

In physics models beyond the Standard Model the couplings between the
Higgs boson and the gauge bosons and fermions can be modified from the
SM theory prediction. In this section we will discuss two such
models, a Higgs portal~\cite{Patt:2006fw} as well as a
strongly-interacting light Higgs~\cite{Espinosa:2010vn}.

\subsection{Higgs Portal}
In the Higgs portal model, an additional hidden sector is added which is
a singlet under the SM gauge groups. A connection to the SM is only
possible via a term connecting the Higgs field of the SM $\Phi_s$ with
that of the hidden sector $\Phi_h$
\begin{equation*}
\mathcal{L} \propto \Phi_s^\dagger \Phi_s \Phi_h^\dagger \Phi_h \ .
\end{equation*}
After electro-weak symmetry-breaking both fields obtain a vev. The two
physical Higgs bosons of the SM and the hidden sector mix and need to be
rotated into mass eigenstates
\begin{equation}
\begin{pmatrix}
H_1 \\ H_2
\end{pmatrix}
= 
\begin{pmatrix}
\cos\chi & \sin\chi \\ -\sin\chi & \cos\chi
\end{pmatrix}
\begin{pmatrix}
H_s \\ H_h
\end{pmatrix} \ .
\end{equation}
The parameter $\cos\chi$ corresponds to our single-parameter modifier
$\Delta_H$ defined before.
The cross sections and branching ratios then change in the following way
from their SM value for $H_1$
\begin{align}
\sigma &= \cos^2 \chi \cdot \sigma^\text{SM} \\
\Gamma_\text{vis} &= \cos^2 \chi \cdot
\Gamma^\text{SM}_\text{vis} \\
\Gamma_\text{inv} &= \cos^2 \chi \cdot
\Gamma^\text{SM}_\text{inv} + \Gamma_\text{hid} \ .
\end{align}
$\Gamma^\text{SM}_\text{inv}$ is induced by Higgs decays into four
neutrinos, which has a negligible rate for light Higgs bosons. The
partial decay width into the hidden sector $\Gamma_\text{hid}$ is a free
parameter and depends on the structure, i.e.\ couplings and masses, of
the hidden sector particles, being zero if they are all heavy.
Corresponding equations hold for $H_2$ with the replacement $\cos\chi
\leftrightarrow \sin\chi$ plus possibly decays $H_2 \rightarrow H_1 H_1$
added, if this channel is kinematically allowed.

\begin{figure}
\centerline{
\includegraphics[width=0.45\textwidth]{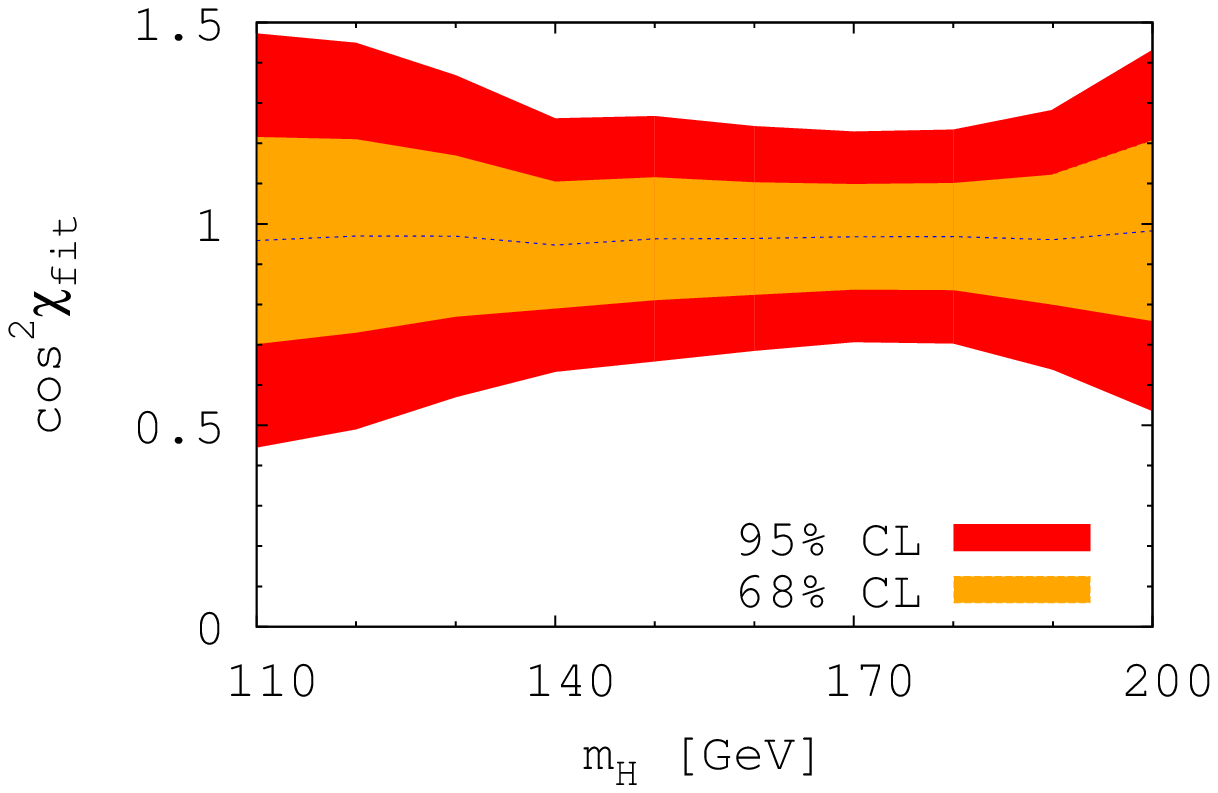}
\quad
\includegraphics[width=0.45\textwidth]{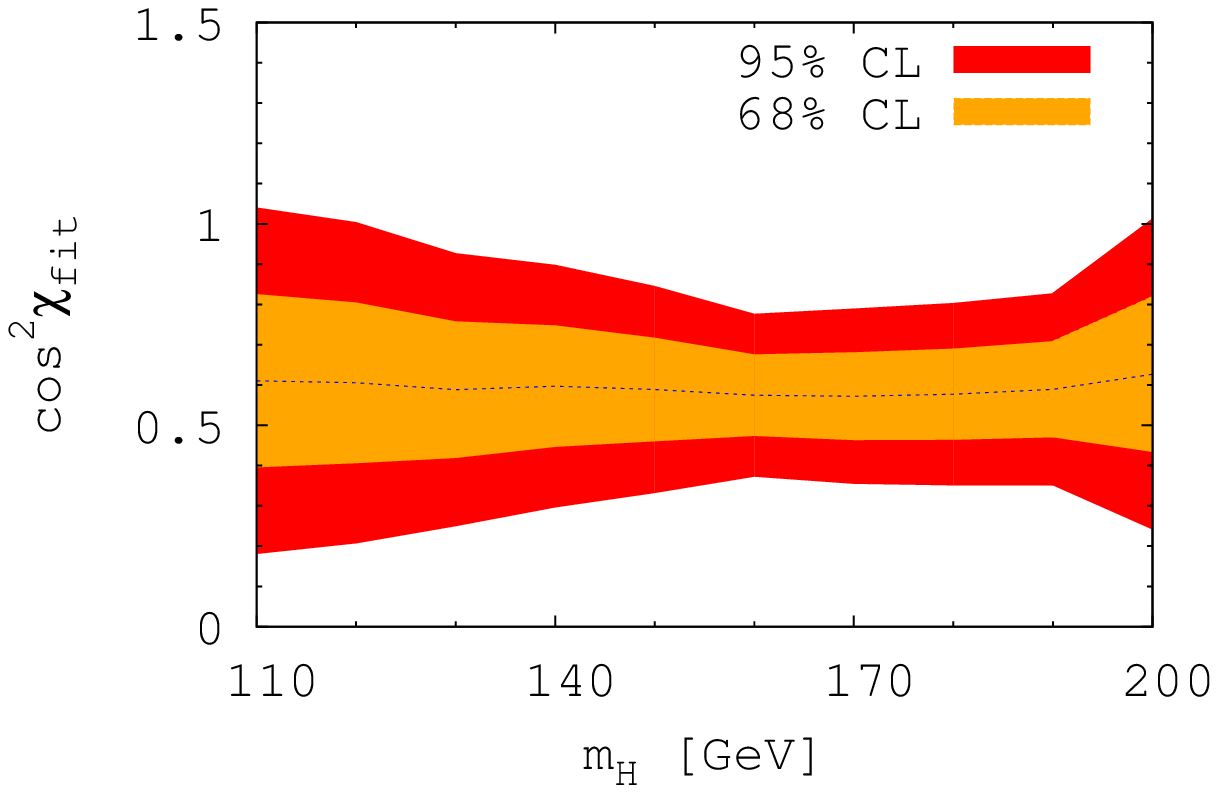}
} 
\caption{Precision in the Higgs portal model assuming a theory input of
$\cos^2\chi=1$ \textit{(left)}, corresponding to the SM value,
and an input value of $\cos^2\chi=0.6$ \textit{(right)}.
Numbers assume LHC data at 14 TeV with an integrated luminosity of 30
fb$^{-1}$ and no invisible decay modes.}
\label{fig:portal1}
\end{figure}
\begin{figure}
\centerline{
\includegraphics[width=0.45\textwidth]{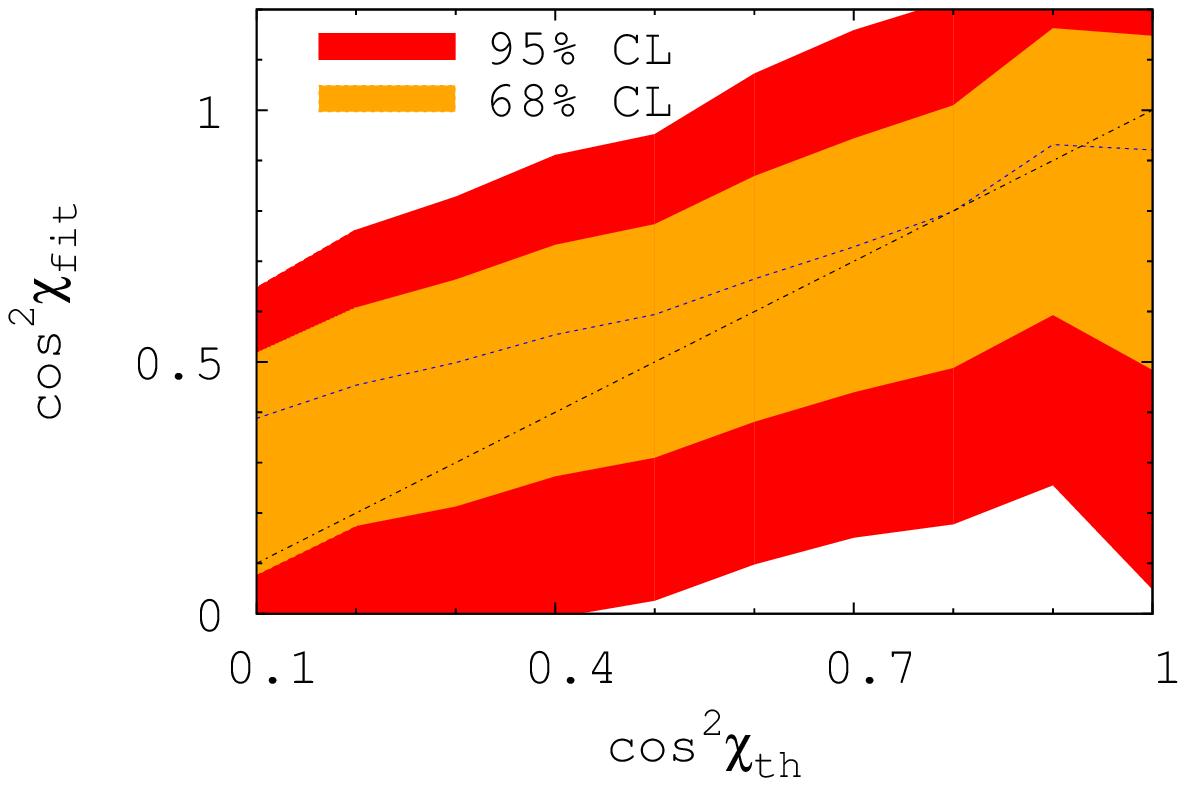}
\quad
\includegraphics[width=0.45\textwidth]{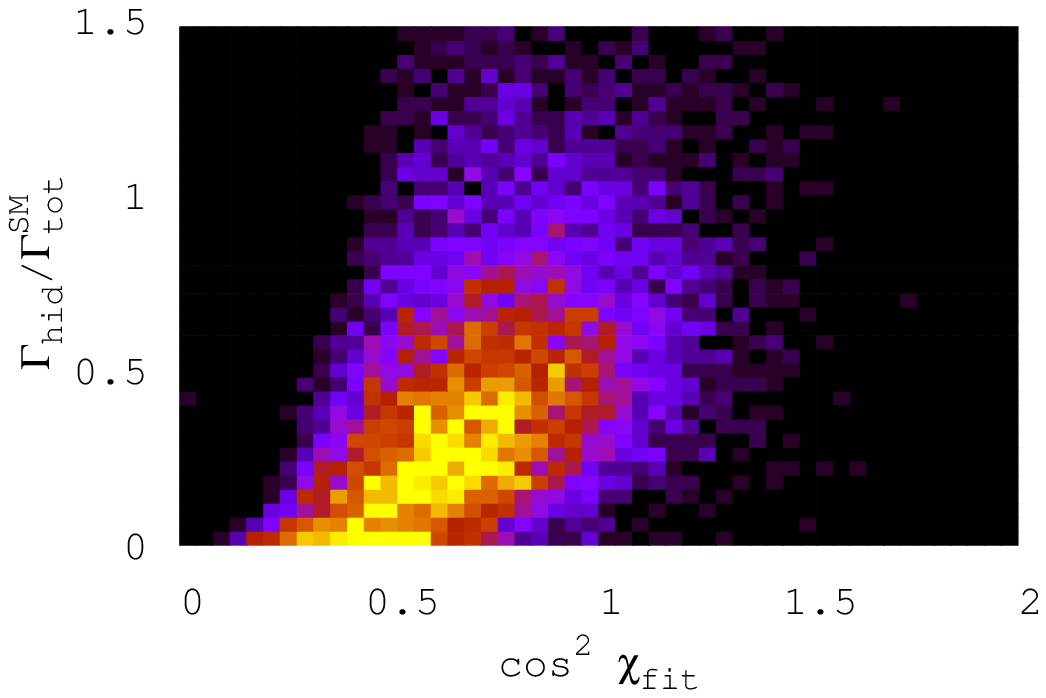}
}
\caption{Precision in the Higgs portal model assuming a Higgs mass of
120 GeV and including invisible decays with $\Gamma_\text{hid} = \sin^2
\chi \cdot \Gamma_\text{tot}^\text{SM}$. We show the fitted value
$\cos^2\chi_\text{fit}$ over the input value $\cos^2\chi_\text{th}$
\textit{(left)}, as well as the correlation between
$\cos^2\chi_\text{fit}$ and $\Gamma_\text{hid}$ for an input value of
$\cos^2\chi_\text{th}=0.6$ \textit{(right)}.  Numbers assume LHC data at
14 TeV with an integrated luminosity of 30 fb$^{-1}$. The brightness in
the correlation plot on the right-hand side denotes the resulting
log-likelihood.}
\label{fig:portal2}
\end{figure}

In Fig.~\ref{fig:portal1} we present the fitted $\cos^2\chi$ as a
function of the Higgs mass in a scenario where the Higgs contains no
additional decay modes into invisible particles. $\cos^2\chi$ is a free
parameter, which is not constrained to its physical range. On the
left-hand side the input value of $\cos^2\chi$ is chosen as one,
corresponding to the SM scenario. Hence, this curve corresponds to the
$\Delta_H$ line of Fig.~\ref{fig:mh14}. The central value is correctly
reproduced by the fit. Errors at the 95\% CL range between 25\% and 50\%
with the highest precision obtainable for a mass of 170 GeV. On the
right-hand side the same plot is shown but now with an input value of
$\cos^2\chi = 0.6$.  The central values are shifted down to smaller
values, but the absolute size of the errors stays approximately the
same. This is due to the fact that most channels have large backgrounds,
which are not affected by a reduction in signal cross section. At the
chosen luminosity of 30 fb$^{-1}$ these give the dominant effect. Also,
the value $\cos^2\chi = 1$ is outside the 95\% CL band over almost
the whole mass range.  Therefore, in this scenario the SM could be
excluded at the 95\% CL.

Figure~\ref{fig:portal2} shows the fitted over the input $\cos^2\chi$
for a Higgs boson mass of 120 GeV. Now decays into the invisible sector
are also included with a partial width of $\sin^2\chi$ times the SM
Higgs width. This corresponds for example to the case where the hidden
sector is an exact copy of the SM sector. Correspondingly, a measurement
of the branching ratio into invisible particles is added~\cite{wbf_inv,
atlas_tdr, cms_tdr}. This will be possible only with a rather low
precision at the LHC. Therefore, the expected accuracy on $\cos^2\chi$
is much lower than in the previous case, as can be seen on the left-hand
side of Fig.~\ref{fig:portal2}.  Also, at low values of $\cos^2\chi$, we
see a deviation of the fitted value, tending to be larger than the input
one. This is because only measurements with a positive signal are taken
into account. Positive fluctuations are hence always included, while
negative ones might get removed. The observation of a Higgs signal
therefore favours larger values of the coupling.  On the right-hand side
of Fig.~\ref{fig:portal2} a correlation plot between the invisible decay
width and the fitted $\cos^2\chi$ is depicted for an input value of
$\cos^2\chi = 0.6$. A strong correlation between the two variables is
visible, which is the origin of the large errors on $\cos^2\chi$
observed before. This correlation is due to the total width of the Higgs
boson, where the invisible decay width enters.  As the denominator in
the branching ratio it enters into all measurements.

\subsection{Strongly-interacting Light Higgs}

In strongly-interacting light Higgs models~\cite{Espinosa:2010vn}, the
Higgs boson emerges as a pseudo-Goldstone boson of a new,
strongly-interacting sector. As a pseudo-Goldstone boson, the Higgs can be
much lighter than the other particles of the theory and therefore be in
the mass range still allowed by all experimental constraints, while the
other ones can be chosen heavy enough to avoid constraints from direct
searches.  Modifications of the Higgs-boson couplings can be
parametrised by $\xi = \left(\frac{v}{f}\right)^2$, where $v = 246
\text{ GeV}$ is the SM Higgs vev and $f$ the Goldstone scale. The limit
$f \rightarrow \infty$ corresponds to the SM, while $ f = v $ are
Technicolour models.

There are two important phenomenological implementations. In the first
one, called MCHM4, all couplings of the Higgs boson to other particles
scale with $\sqrt{1-\xi}$. Therefore, the results of the Higgs portal in
the previous subsection can be reused by identifying $\cos^2\chi =
1-\xi$ and setting invisible decay modes to zero.
In the second one, MCHM5, the couplings change differently for vector
bosons and fermions
\begin{align*}
g_{VVH} &= g_{VVH}^\text{SM} \cdot \sqrt{1-\xi} \\
g_{f\bar{f}H} &= g_{f\bar{f}H}^\text{SM} \cdot \frac{1-2\xi}{\sqrt{1-\xi}} \ .
\end{align*}
The latter one has the interesting feature that the coupling vanishes
for $\xi=0.5$ and flips its sign for values below that.
These models also show significant deviations in Higgs pair-production
processes~\cite{Grober:2010yv}, which we will not consider further here.

\begin{figure}
\centerline{
\includegraphics[width=0.45\textwidth]{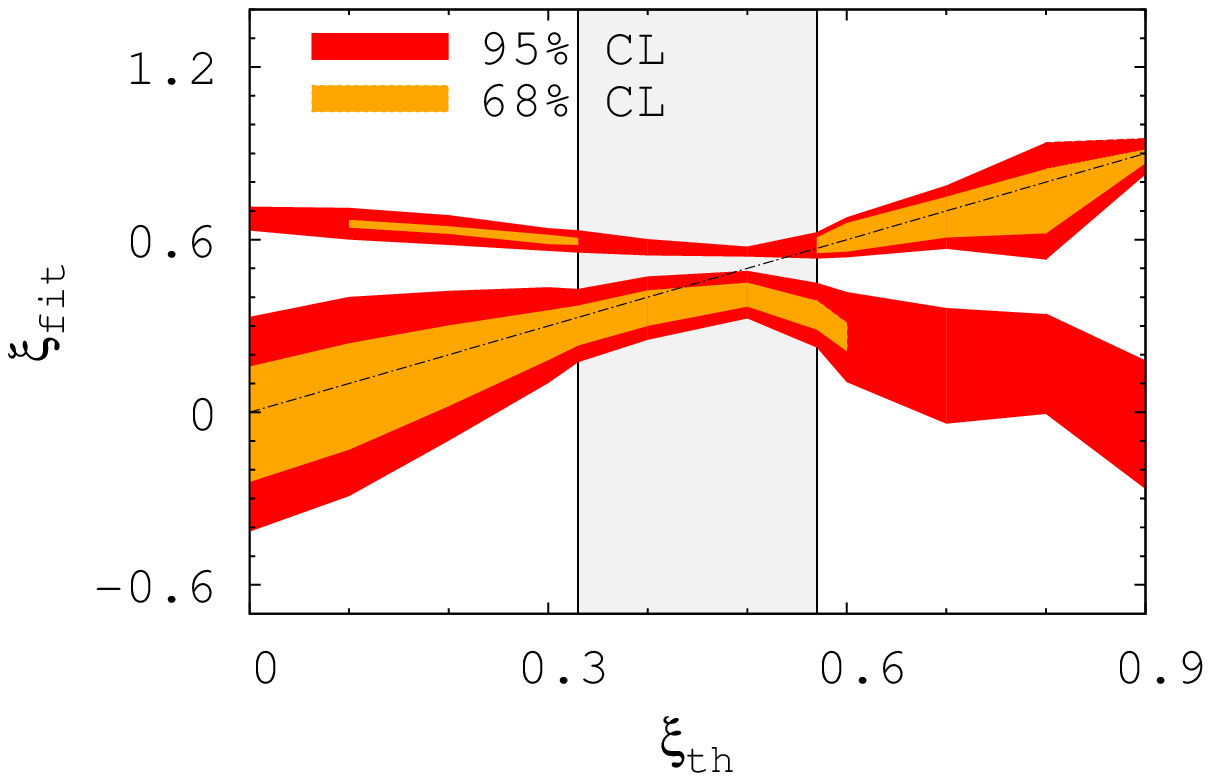}
\quad
\includegraphics[width=0.45\textwidth]{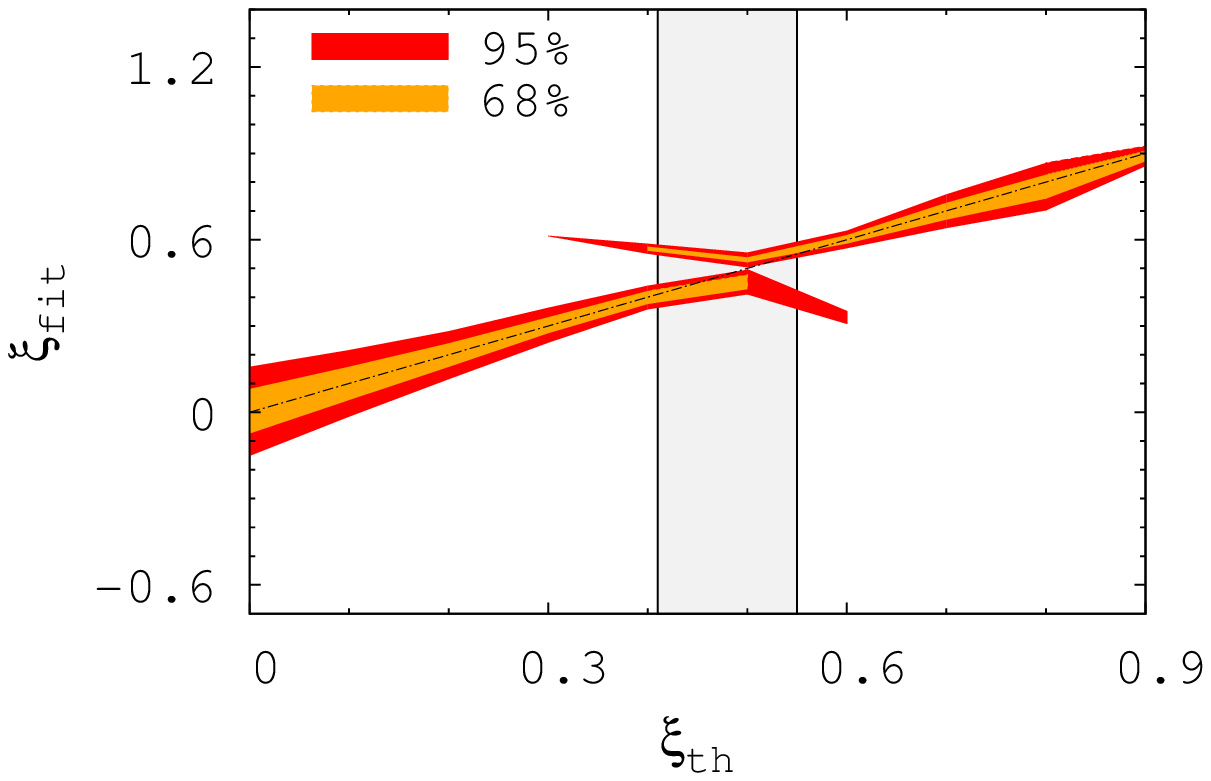}
} 
\caption{Best-fit values and 68\% and 95\% CL error bands  in the MCHM5
model for the LHC at 14 TeV assuming a Higgs boson mass of 120 GeV as
function of the input value $\xi_\text{th}$. Results are shown for an
integrated luminosity of 30 \textit{(left)} and 300
fb$^{-1}$ \textit{(right)}.}
\label{fig:silh1}
\end{figure}
\begin{figure}
\centerline{
\includegraphics[width=0.45\textwidth]{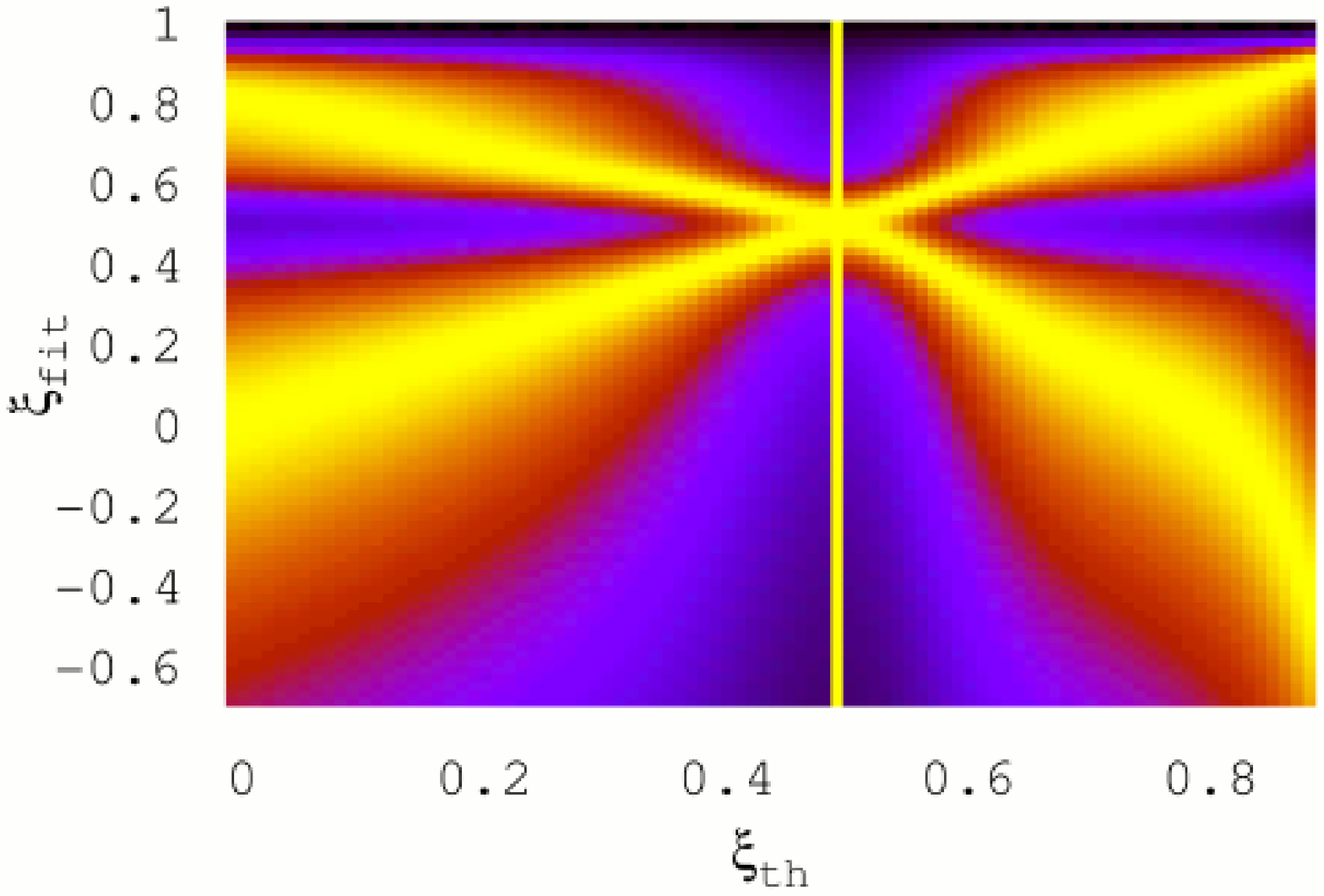}
\quad
\includegraphics[width=0.45\textwidth]{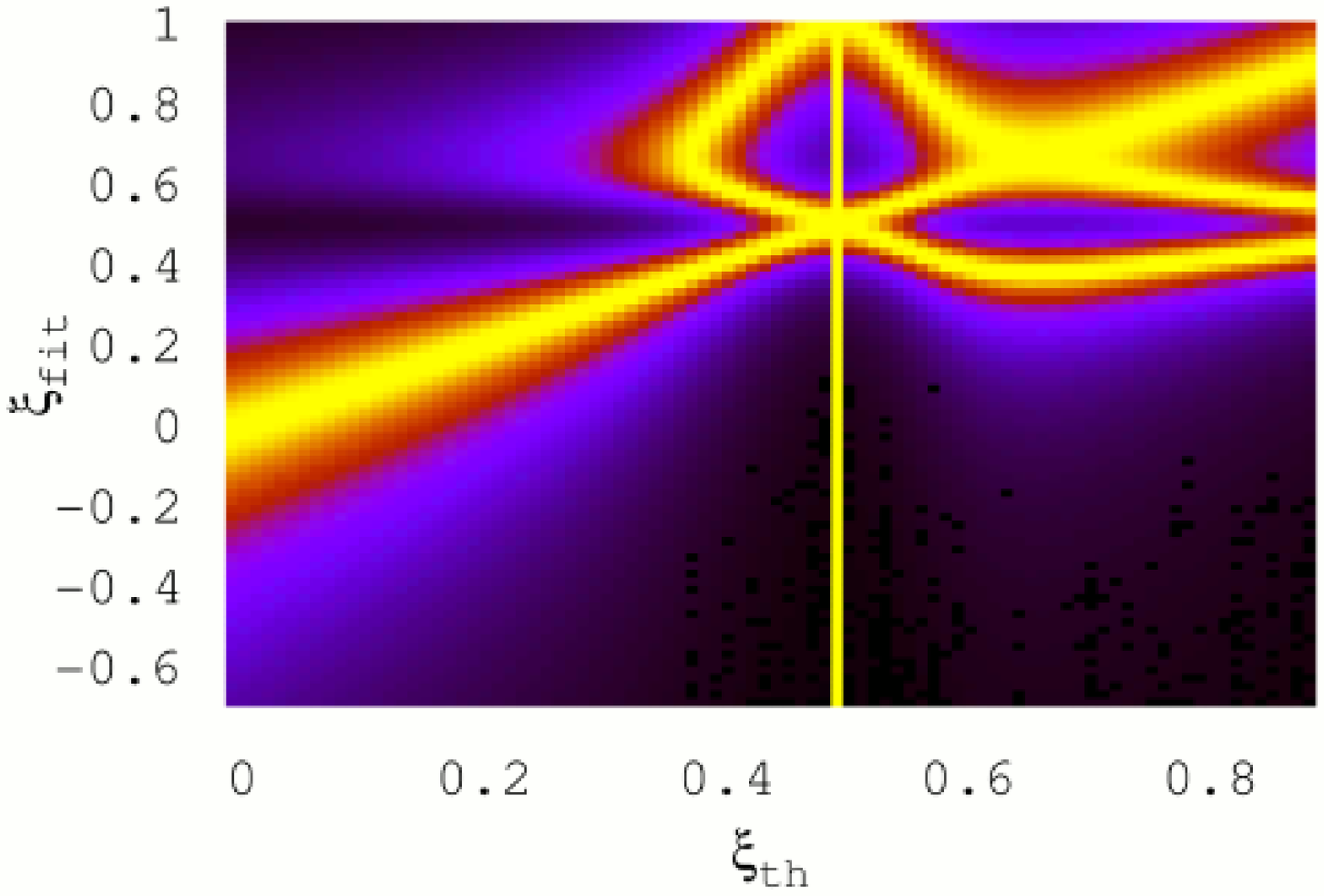}
}
\caption{Best-fit distribution for two main channels:
gluon-fusion production with decay into photons \textit{(left)} and
vector-boson-associated production with decay into bottom quarks
\textit{(right)}.
Results are shown for a Higgs boson of 120 GeV at the LHC at 14 TeV and
an integrated luminosity of 30 fb$^{-1}$. The brightness denotes the
resulting log-likelihood.
}
\label{fig:silh2}
\end{figure}

In Fig.~\ref{fig:silh1} we depict the fitted value of $\xi$ over the
input one for an integrated luminosity of 30 (left) and 300 fb$^{-1}$
(right) at the 14 TeV LHC. The shaded region around $\xi=0.5$ denotes
the region where the cross sections are so low, that with the given
luminosity no evidence of a Higgs boson is yet expected.
For the lower luminosity there are always two possible solutions. One
corresponds to the correct solution, while the other originates from the
ambiguity in the fermion-Higgs coupling. The sign of the coupling is
only observable as interference between $W$-boson and top-quark loop in
the effective photon coupling. With the higher luminosity this
degeneracy is lifted, as can be seen on the right-hand side of
Fig.~\ref{fig:silh1}. 
This is further demonstrated in Fig.~\ref{fig:silh2}. Here we show the
log-likelihood for 30 fb$^{-1}$ in the two individual channels which
contribute most to the parameter determination. Both channels vanish at
$\xi=0.5$ and therefore for this value the log-likelihood is constant
independent of the parameter. 
The left channel is gluon-fusion Higgs production with decays into photons. 
For each input value two different solutions can be found that cannot
be distinguished, as they yield the same rate. On the right-hand side,
we show the combination of vector-boson associated production channels
with decay into bottom quarks via subjet techniques, which are all
governed by the same coupling factors. Here for $\xi \lesssim 0.4$ only
a single solution exists, while for larger values additional
solutions appear. These do not coincide with the secondary solution of
the first channel, however, so for the nominal values the solution
becomes unique. Therefore, the degeneracy is not a true one, but induced
by fluctuations due to errors and can be lifted with more data.

\section{Conclusions}

The determination of the Higgs-boson couplings is an important task to
verify our understanding of electro-weak symmetry breaking and the Higgs
mechanism. We have studied how well we can measure these couplings at
the LHC and what remains as a task for a future linear collider.
To be independent of any specific new-physics model, we take as a model
the Standard Model, where all Higgs couplings are left as free
parameters. Using the SFitter framework, all experimental and theory
errors, as well as their correlations, can be fully taken into account.
For a single parameter modifying all couplings an error of 9\% is
achievable for a Higgs boson mass of 125 GeV.
We have also interpreted our results in terms of new-physics models,
namely a Higgs portal and a strongly-interacting light Higgs.  For the
former, invisible decay modes provide an additional experimental
challenge. In the latter case, statistical fluctuations lead to
secondary solutions, which also need to be considered.



\section*{Acknowledgments}
We would like to thank the organisers of the LC-Forum for the friendly
atmosphere during the workshop and the possibility to present our
results.
Support by the Deutsche Forschungsgemeinschaft via the
Sonderforschungsbereich/Transregio SFB/TR-9 ``Computational Particle
Physics'' and the Initiative and Networking Fund of the Helmholtz
Association, contract HA-101(``Physics at the Terascale'') is
acknowledged.


\begin{footnotesize}

\end{footnotesize}


\end{document}